\documentclass[notitlepage,aip,reprint,onecolumn]{revtex4-1}

\usepackage{amsmath}
\usepackage{amssymb}
\usepackage{bm}
\DeclareMathSymbol{\hbar}{\mathord}{AMSb}{"7E}
\usepackage{mathptmx}
\usepackage{braket}
\usepackage{mathrsfs}

\usepackage{graphicx}
\usepackage{orcidlink}

\newcommand{\tr}{\mathrm{tr}}
\newcommand{\parprod}{\alpha}
\newcommand{\parsum}{\beta}

\begin{document}


\title{Algebraic structures of the Lindblad equation}

\author{Leonel Bixano \orcidlink{0009-0009-5858-6466}}
\affiliation{Departamento de F\'{\i}sica,
Centro de Investigaci\'on y de Estudios Avanzados del
Instituto Politécnico Nacional,
Av. Instituto Politécnico Nacional 2508,
San Pedro Zacatenco, M\'exico 07360, CDMX.}
\author{Guillermo López-Alvarez}
\author{Victor Alberto Cruz-Barriguete}
\author{V. G. Ibarra-Sierra\orcidlink{0000-0003-2416-3261}}
\author{José Luis Cardoso\orcidlink{0000-0003-3060-1168}}
\author{Juan Carlos Sandoval-Santana\orcidlink{0000-0002-1166-4529}}
\author{Alejandro Kunold\orcidlink{0000-0003-1545-8017}}
\email{akb@azc.uam.mx}
\affiliation{\'Area de F\'isica Te\'orica y Materia Condensada,
Universidad Aut\'onoma Metropolitana,
Azcapotzalco, Av. San Pablo Xalpa 180,
CDMX, 02200, M\'exico}


\begin{abstract}
We investigate the algebraic structure underlying the Lindblad equation
for finite-dimensional open quantum systems. By introducing a suitable
operator representation of the Liouville superoperator,
we show that the dynamics can be formulated in terms of a closed algebra
of Hermitian operators that is independent of the particular physical model.
This formulation reveals that dissipative dynamics requires a substantially
richer algebraic structure than purely unitary evolution, thereby providing
a clear characterization of the additional complexity introduced by the Lindbladian.
The resulting framework naturally leads to parametrizations of the dynamical
map and to differential equations governing its evolution.
We further derive recursion relations that enable the efficient
construction of the algebra for systems of increasing dimension.
Because the algebraic basis is universal, while all model-dependent
information enters through a single set of coefficients,
the proposed approach significantly reduces the computational
cost of constructing the Liouville superoperator compared with direct methods.
To facilitate the implementation of the method, we provide a Mathematica notebook
containing a one-qubit example that can be systematically extended to
an arbitrary number of qubits. The proposed framework therefore provides
both a general mathematical description of finite-dimensional Lindblad
dynamics and a practical foundation for efficient analytical
and numerical implementations.
\end{abstract}

\maketitle

\section{Introduction}

At present, Lindblad master equations play a central role in quantum
science and technology. Although introductory quantum mechanics often
treats physical systems as perfectly isolated, real-world quantum
systems are inevitably noisy and continuously interact with their
surrounding environment \cite{breuer321,manzano2020}.
A complete description of the combined system and environment
is generally intractable because the environment typically
possesses an enormous number of degrees of freedom.
The master-equation approach overcomes this difficulty
by providing an effective description of the reduced dynamics
of the system while incorporating the influence of the environment
through dissipative processes acting within the
system Hilbert space \cite{breuer321,manzano2020}.

The Lindblad equation plays a key role in a wide range
of physical systems, including atomic systems, superconducting transmons
for quantum computing, and spin dynamics in solid-state systems.
In the following, we present a few representative examples.
In quantum information science, the Lindblad equation is widely used to describe
qubit decoherence arising from interactions between the quantum hardware and
its environment, leading to energy relaxation and dephasing
\cite{Krantz2019,Khan2024}.
Within this framework, it provides a natural description of quantum noise
channels such as bit flips, phase flips, and depolarization.
These models are fundamental for the design and characterization of
noisy intermediate-scale quantum (NISQ) processors, the evaluation of
circuit fidelities, and the development of fault-tolerant quantum
error-correction schemes.
Circuit quantum electrodynamics (circuit QED) is a fundamental
building block of superconducting transmon quantum computers
\cite{Kim2023,GoogleQuantumAI2023,GoogleQuantumAI2025,Bravyi2024}
and of ultrasensitive quantum sensing schemes \cite{Degen2017}.
In cavity quantum electrodynamics, the Lindblad equation
accounts for photon leakage through the cavity mirrors,
together with spontaneous emission and other dissipative
channels arising from the coupling to the electromagnetic
environment \cite{Barlow08122015}.
In this context, the Lindblad equation enables the prediction
of steady-state photon emission rates and the transmission
spectra of driven-dissipative cavities.
One of the most remarkable discoveries in quantum biology
and molecular physics is that biological light-harvesting
systems, such as the Fenna--Matthews--Olson complex in green
sulfur bacteria, exploit quantum coherence to transport
energy with a near-$99\%$ efficiency \cite{Mohseni2008}.
In this context, the Lindblad equation provides a framework
for describing how excitons propagate through a molecular
network while interacting with a warm, noisy protein
environment.
This model revealed the counterintuitive phenomenon of
environment-assisted quantum transport (also known as
dephasing-assisted transport), demonstrating that a carefully
tuned level of environmental noise can actually prevent
energy from becoming trapped, thereby optimizing biological
energy conversion.
Traditional thermodynamics breaks down at the atomic scale,
where quantum fluctuations become significant.
Quantum thermodynamics extends the concepts of heat,
work, and entropy to microscopic open quantum systems
\cite{Picatoste2024}. By treating thermal baths as
environmental reservoirs, the Lindblad framework is used
to model the performance of quantum heat engines,
refrigerators, and multiqubit thermal machines
\cite{Caselli2026}. It enables the calculation of work
extraction rates, heat currents, and entropy production
while remaining consistent with the Second Law of
Thermodynamics.
In condensed-matter systems, the Lindblad equation is widely
used to model dissipative transport, spin relaxation, exciton dynamics,
and driven-dissipative lattice systems, where electrons or spins
propagate through complex crystalline structures while interacting
with localized defects and phonons \cite{manzano2020}. Within this
framework, it enables the study of nonequilibrium transport,
particle currents, and the evolution of localized quantum states in
the presence of environmental reservoirs, providing deeper insights
into transport phenomena, topological protection, and dissipative
phase transitions.
In semiconductor spintronic systems, where Mn atoms are embedded
in II--VI semiconductor self-assembled quantum dots, the spin of
the optically created electron-hole pair interacts with the
electrons of the Mn atom \cite{Jamet2013}. The strong mixing
between the optical excitation and the Mn spin affects the spin
state of the Mn through the exchange interaction, providing a
mechanism for optical control \cite{Urbaszek2013}. In this case,
the complexity arising from the interaction between the Mn quantum
levels and the electromagnetic reservoir is greatly reduced by the
Lindblad equation, whose Lindbladian accounts for the radiative
decay of the exciton (through its irreversible coupling to the
photon modes) and the relaxation of the Mn spin (through its
irreversible coupling to the phonon modes).
Atomic ensembles constitute an ideal platform for studying
cooperative light-matter interactions.
In these systems, emitters interact through the electromagnetic
field via coherent dipole-dipole interactions and collective
radiative decay.
The Lindblad equation naturally incorporates both coherent
and dissipative couplings, enabling the description of
collective emission, superradiance, subradiance, and
nonlinear optical phenomena in driven atomic systems
\cite{Reitz2022,Yanes2025}.

Despite their broad range of applications, Lindblad master equations
are almost always solved numerically by constructing the Liouville
superoperator directly in a chosen basis.
This procedure rapidly becomes computationally demanding as the Hilbert-space
dimension increases and obscures the underlying algebraic structure of the dynamics.
In this work, we develop a general algebraic formulation of the Lindblad equation
that separates the universal basis-dependent structure from the model-dependent
physical information, leading to a compact representation of the Liouville
superoperator together with an efficient computational framework.
To facilitate the implementation of the proposed approach,
we provide a Mathematica \cite{Mathematica} notebook
through GitHub \cite{lindbladalgebra}
containing illustrative
examples of the algebraic method.
The notebook is designed to be readily extended to systems
with an arbitrary number of qubits and serves as a practical
starting point for applications to more complex open quantum
systems.

The paper is organized as follows.
Section~\ref{sec:liouvilleequation} introduces
the Lindblad equation and the two contributions to the
Liouville superoperator.
Section~\ref{sec:liouvillespace} introduces
the Liouville space.
The techniques required to vectorize the density matrix
in terms of a Hermitian matrix basis, thereby giving rise
to the coherent vector representation, are developed in
Section~\ref{sec:vectorization}.
There, the Liouville superoperator associated with the
Lindblad equation is expressed as a linear combination
of the elements of an algebra intimately related to the
Lindbladian.
The algebraic structure of these elements, together with
their orthogonality properties, is established in
Sections~\ref{sec:algebraicstructure}
and~\ref{sec:innerproductspaceX}.
Section~\ref{sec:algebraicstructure} derives the algebraic
relations satisfied by the structure-constant superoperators
appearing in the Liouville superoperator.
Section~\ref{sec:innerproductspaceX} shows that,
under certain conditions, this set
forms an inner product space
whose elements are orthogonal with respect to the
Hilbert-Schmidt inner product.
This property greatly simplifies the parametrization
of the evolution operator.
Finally, Section~\ref{sec:dynamicalmap}
combines the algebraic, vector-space,
and inner-product structures of
$\mathfrak{X}_{n^2}$ to derive explicit expressions
for the Liouville superoperator and the associated
dynamical map.
Section~\ref{seq:examples}
presents the one-qubit case as an illustrative example.
We explicitly construct the matrix basis, compute the
corresponding structure-constant superoperators,
and express the Liouville superoperator in terms of them.
We then verify that the Liouville superoperator and the
differential equations derived from the algebraic
formulation are fully equivalent to those obtained
through the direct approach.
Section~\ref{sec:conclusions}
contains our concluding remarks.

\section{Liouville equation}\label{sec:liouvilleequation}
The dynamics of open quantum systems, i.e., systems that interact
with an environment, are typically described by Markovian master equations
\cite{breuer321,carmichael1993open}.
Under rather general physical assumptions, the Lindblad equation
provides the most general form of a continuous-time Markovian
quantum evolution
\cite{gorini1976,lindblad1976generators,chruscinski2017,manzano2020}.
It describes the temporal evolution of the reduced density matrix
$\rho(t)$ of a quantum system interacting with an environment,
and possibly external fields, through an equation of the form
\begin{equation}
\frac{d}{dt}\rho(t)=L\rho(t),
\label{eq:Lindblad01}
\end{equation}
where $L$ is the Liouvillian superoperator.
In this context, a superoperator is a linear mapping acting
on the vector space of linear operators, such as the density matrix
$\rho(t)$ itself.

The Liouvillian superoperator can be decomposed into two contributions,
\begin{equation}
L = L_H + L_L.
\label{eq:liouville01}
\end{equation}
The Hamiltonian contribution is
\begin{equation}
L_H\rho(t)= -\frac{i}{\hbar}[H,\rho(t)],
\label{eq:liouvilleHpart00}
\end{equation}
with $H$ an effective Hamiltonian of the open system.
In general, this Hamiltonian differs from the bare Hamiltonian
of the corresponding closed system due to the interaction with
the environment \cite{breuer321,davies1976quantum}.
A well-known example is the Lamb shift \cite{lamb1947fine},
which corresponds to an environment-induced shift of the system
energy levels.

The dissipative contribution is given by
\begin{equation}
L_L\rho(t)=
\frac{1}{2}\sum_{i,j}\gamma_{i,j}
\left[
2A_j\rho(t)A_i^\dagger
-\{A_i^\dagger A_j,\rho(t)\}
\right],\label{eq:liouvilleLpart00}
\end{equation}
where $\{A,B\}=AB+BA$ denotes the anticommutator,
$\gamma_{i,j}$ are the transition rates and
$A_i$ are the corresponding jump operators associated with decoherence
and dissipative processes that satisfy with the Hilbert-Schmidt
orthonormality condition\cite{KLendi_1987,breuer321}
\begin{equation}
\tr\left[A_i^\dagger A_j\right]=\delta_{i,j}
\end{equation}
In the particular case in which $\gamma_{i,j}=0$ for all $i,j$,
the Lindblad equation reduces to the von Neumann equation
\begin{equation}
\frac{d}{dt}\rho(t)=L_H\rho(t).
\label{eq:vonNeumann01}
\end{equation}

The Liouville superoperator \(L\) is the infinitesimal
generator of a completely positive quantum dynamical
semigroup \cite{KLendi_1987}. Such a semigroup is a
one-parameter family of completely positive, trace-preserving
linear maps \(V(t)\), known as quantum dynamical maps
(or propagators), that describe the continuous-time evolution
of an open quantum system under
Markovian dynamics \cite{havel2003}.
These maps satisfy the semigroup composition law
\(V(t)V(s)=V(t+s).\)
The action of the propagator on an initial state \(\rho(0)\)
therefore yields the density matrix at time \(t\),
\begin{equation}
\rho(t)=V(t)\rho(0).
\label{eq:dynamicalmap00}
\end{equation}
A general representation of a
quantum dynamical semigroup
is provided by the Kraus operator-sum
decomposition \cite{kraus1971}
\begin{equation}
\rho(t)=V(t)\rho(0)=\sum_l K_l(t)\rho(0)K_l^\dagger(t),
\end{equation}
where the operators $K_l(t)$ are known as Kraus operators, and the
number of Kraus operators satisfies $l \leq n = m^2$, where $m$ is the
dimension of the Hilbert space on which $\rho(t)$ acts.

\section{Liouville space}
\label{sec:liouvillespace}
The Hilbert space associated with an $m$-level quantum system,
denoted by $\mathcal{H}_m$, is spanned by the orthonormal states
$\{\ket{1}, \ket{2}, \dots, \ket{m}\}$.
These states induce an orthonormal operator basis
with respect to the Hilbert--Schmidt inner product
\begin{equation}
\left\langle A, B \right\rangle = \tr\left[A^\dagger B\right].
\end{equation}
The operators forming this basis are elements of the larger
Hilbert space
$\mathcal{L}_{n}=\mathcal{H}_m\otimes\mathcal{H}_m^*$,
of dimension $n=m^2$, where $*$ denotes the dual space.
This space is commonly referred to as the Liouville or von Neumann space.

Usually, to vectorize the Liouville equation, the matrix-unit basis
\begin{multline}
\mathfrak{e}_n=\left\{e_1, e_2, \dots, e_n\right\}\\
=\left\{\ket{1}\bra{1}, \ket{2}\bra{1}, \dots, \ket{m}\bra{1},\dots,
\ket{1}\bra{2}, \ket{2}\bra{2}, \dots, \ket{m}\bra{2},\dots,
\ket{1}\bra{m}, \ket{2}\bra{m}, \dots, \ket{m}\bra{m}
\right\},\label{eq:matrixunitbasis01}
\end{multline}
is chosen
\cite{dirac1981principles, Hioe1981, horn2012matrix, Kunold_2024},
which satisfies
\begin{equation}
\left\langle e_i, e_j \right\rangle
=\tr\left[e_i^\dagger e_j\right]
=\delta_{i,j}.
\end{equation}

However, many aspects of the vectorization process can be
considerably simplified by choosing instead
an orthonormal basis of Hermitian operators,
$\mathfrak{h}_n=\{h_1,h_2,\dots,h_n\}$ that satisfies
\begin{equation}
\left\langle h_i, h_j \right\rangle
=\langle h_i|h_j\rangle
=\tr\left[h_i h_j\right]
=\delta_{i,j},
\label{eq:orthonormh00}
\end{equation}
where $\ket{h_i}$ denotes the vectorized form of the operator $h_i$ in Liouville space.

Two such simplifications are that the expansion coefficients
of a Hermitian operator in this basis are real as well as the matrix elements
of the Liouville linear maps.
Working with purely real quantities offers significant computational advantages,
especially because it allows the resulting differential equations
to be written entirely in real form.

Any operator $O$ acting on $\mathcal{H}_m$ can be expanded as
a linear combination of the elements of $\mathfrak{h}_n$ as
\begin{equation}
O=\sum_i \tr\left[O h_i\right] h_i,
\label{eq:expansionA00}
\end{equation}
provided that $O\in \mathcal{L}_n$.

In particular, the density matrix can be written as
\begin{equation}
\rho(t)=\sum_i \rho_i(t)\, h_i,
\label{eq:rhoexpansion00}
\end{equation}
where the expansion coefficients are
\begin{equation}
\rho_i(t)
=\tr\left[h_i \rho(t)\right]
=\langle h_i | \boldsymbol{\rho}(t)\rangle.
\end{equation}
The vector formed by these coefficients,
\begin{equation}
\boldsymbol{\rho}(t)=
\left(\rho_1(t),\rho_2(t),\dots,\rho_n(t)\right)^{\top},
\end{equation}
is referred to as the coherence vector \cite{KLendi_1987}.

Any superoperator or linear map, such as $L_H$, $L_L$, or
the complete Liouvillian $L$,
acting on $\rho(t)$ can therefore be represented as a matrix acting on Liouville space
$\mathcal{L}_n$.
The matrix elements of a superoperator $L$ in this basis are given by
\begin{equation}
\left(L\right)_{i,j}
=\tr\left[h_i\, L h_j\right].
\label{eq:orthonormalh00}
\end{equation}

\section{Vectorization of the Lindblad equation}
\label{sec:vectorization}
\label{sec:vectorization}
Usually, the Lindblad equation is vectorized
by stacking matrix elements into a vector representation
\cite{dirac1981principles, Hioe1981, horn2012matrix, Kunold_2024},
\begin{equation}
\boldsymbol{\rho}(t)=\left(
\rho_{1,1}(t), \rho_{2,1}(t),\dots, \rho_{m,1}(t), \dots
\rho_{1,2}(t), \rho_{2,2}(t),\dots, \rho_{m,2}(t), \dots
\rho_{1,m}(t), \rho_{2,m}(t),\dots, \rho_{m,m}(t)
\right)^\top .
\end{equation}
This is equivalent to projecting the density matrix
onto the base in Eq. \eqref{eq:matrixunitbasis01}\cite{Kunold_2024}.
Proceeding in this manner yields the following matrix representation
for the Hamiltonian part
\cite{havel2003, Ramusat2021quantumalgorithm, Kamakari2022}
\begin{equation}
\left(L_H\right)_{i,j}=-\frac{i}{\hbar}\left(
I\otimes H-H^\top \otimes I
\right)_{i,j},
\label{eq:LHelementary}
\end{equation}
and the Lindbladian part
\begin{equation}
\left(L_L\right)_{i,j}=\frac{1}{2}\sum_{k,l}\gamma_{k,l}
\left(
2 A_k^* \otimes A_l 
- I\otimes A_k^\dagger A_l - A_k^* A_l^\top \otimes I 
\right)_{i,j},
\label{eq:LLelementary}
\end{equation}
of the Lindblad linear map
where $I$ denotes the identity operator.
However, as mentioned earlier, this procedure generally leads
to complex components of the coherence vector, as well as
complex matrix elements of the Liouville superoperator
\((L)_{i,j}\).
An alternative approach consists of projecting the Liouville
linear map onto a basis of matrices that is orthonormal under
the Hilbert-Schmidt inner product
\cite{Bloch1946,Fano1957,Hioe1981,alicki2007quantum}.
In the context of nuclear magnetic resonance,
this approach is commonly referred to as the direct method.
Although this method is more general, it may also yield
complex coherence vectors and Liouville superoperators.
As discussed previously, Eqs.~\eqref{eq:LHelementary}
and \eqref{eq:LLelementary} correspond to the particular case
in which the direct method is applied using
the matrix-unit basis defined in
Eq.~\eqref{eq:matrixunitbasis01}.
A more convenient representation can be obtained by choosing
instead a Hermitian matrix basis \(\mathfrak{h}_n\),
which can be selected so that both the coherence vector
and the Liouville superoperator have real components.

We begin by considering the Hamiltonian contribution to the Liouville
superoperator.
To this end, we expand the Hamiltonian in terms
of the elements of $\mathfrak{h}_n$ as
\begin{equation}
H = \sum_j H_j h_j,
\qquad
H_j = \tr\!\left[h_j H\right].
\label{eq:hamiltonianexpansion00}
\end{equation}
It is convenient to define the Hamiltonian vector as
\begin{equation}
\boldsymbol{H}=(H_1, H_2, \dots, H_n)^\top.
\label{eq:hamiltonianvector00}
\end{equation}
Substituting the expansions of the Hamiltonian
and the density matrix,
given in Eqs.~\eqref{eq:hamiltonianexpansion00}
and \eqref{eq:rhoexpansion00}, respectively,
into the Hamiltonian contribution to the
Liouville linear map, Eq.~\eqref{eq:liouvilleHpart00},
requires the commutation relations
\begin{equation}
\left[h_j,h_k\right]
=
i\sum_i c_{j,k,i}\,h_i,
\label{eq:struconst00}
\end{equation}
where $c_{j,k,i}$ are the Lie structure constants.
The structure constants \(c_{i,j,k}\), which encode the algebraic
properties of the basis \(\mathfrak{h}_n\), can be expressed
in terms of its elements as
\begin{equation}
c_{i,j,k}
=\frac{1}{i}\tr\left[h_i\left[h_j, h_k\right]\right].
\label{eq:struconst01}
\end{equation}
These constants possess several useful properties,
stemming from the trace and commutator operations,
that will prove useful in what follows.
In particular, they are real-valued,
completely antisymmetric under the permutation of any two indices,
\begin{equation}
c_{i,j,k} = -c_{j,i,k} = -c_{i,k,j},
\end{equation}
and therefore invariant under cyclic permutations of the indices,
\begin{equation}
c_{i,j,k} = c_{k,i,j} = c_{j,k,i}.
\end{equation}

Substituting the commutator in Eq.~\eqref{eq:liouvilleHpart00}
by its expansion in terms of the structure constants,
Eq.~\eqref{eq:struconst00}, the Hamiltonian contribution
to the Liouville superoperator can be written as
\cite{Kunold_2024}
\begin{equation}
L_H\rho(t)
=\frac{1}{\hbar}\sum_{i,j,k} H_j\, h_i\, c_{i,j,k}\,\rho_k(t).
\end{equation}
Projecting this expression onto the orthonormal basis
\(\mathfrak{h}_n\) using the orthonormality condition,
Eq.~\eqref{eq:orthonormalh00}, yields the matrix elements
of the Hamiltonian part of the Liouville superoperator,
\begin{equation}
\left(L_H\right)_{i,j}
=\frac{1}{\hbar}\sum_{k} H_k\, c_{k,i,j}.
\end{equation}
Introducing the superoperators associated with the structure constants,
defined by
\begin{equation}
(C_k)_{i,j}=c_{k,i,j},
\label{eq:struconsc00}
\end{equation}
the Hamiltonian contribution to the Liouville superoperator
acquires the compact form
\begin{equation}
L_H=\frac{1}{\hbar}\sum_{k} H_k\, C_k.
\end{equation}

We now turn to the projection of the Lindbladian contribution
to the Liouville operator onto the elements of $\mathfrak{h}_k$.
Whereas the Hamiltonian part is governed by the commutator
and thus naturally gives rise to the structure constants
of the Lie algebra, the Lindbladian contribution involves
anticommutator terms, leading to an additional algebraic structure.
To account for this, we introduce a second set of structure constants
associated with the anticommutator of the basis elements
\cite{PhysRevE.100.053305,KLendi_1987},
\begin{equation}
\left\{h_i, h_j\right\} = \sum_k b_{i,j,k} h_k.
\end{equation}
The coefficients $b_{i,j,k}$ define the structure constants
of the Jordan algebra associated with the matrix algebra
of $\mathfrak{h}_n$.
They can be explicitly calculated using
the Hilbert-Schmidt inner product as
\begin{equation}
b_{i,j,k}=\tr\left[h_i\left\{h_j, h_k\right\}\right].
\end{equation}
Using the identities of the trace and the anticommutator,
it can be shown that these constants are real-valued and
completely symmetric
under permutations of any two indices, i.e.,
\begin{equation}
b_{i,j,k} = b_{j,i,k} = b_{i,k,j} = b_{k,j,i}.
\end{equation}
Although the Lindbladian contains anticommutator terms,
its structure is considerably more intricate,
giving rise to contributions beyond those simply proportional to \(b_{i,j,k}\).
To identify these contributions, we expand the jump operators \(A_a\) as
\begin{equation}
A_i = \sum_k \tr\left[A_i h_k\right] h_k = \sum_k w_{i,k} h_k,
\end{equation}
where \(w_{i,k}=\tr\left[A_i h_k\right]\),
and substitute this expansion into the Lindbladian
contribution to the Liouville operator, \(L_L\),
\begin{equation}
L_L\rho(t)=
\frac{1}{2}\sum_{j,k,l}\Gamma_{k,l}
\left[
2h_l h_j h_k
-\{h_k h_l,h_j\}
\right]\rho_k(t),
\label{eq:liouvilleLpart01}
\end{equation}
where
\begin{equation}
\Gamma_{k,l} = \sum_{i,j} w_{i,k}^* w_{j,l}\gamma_{i,j}.
\label{eq:Gammafromgamma}
\end{equation}
Projecting $L_L\rho(t)$ on to the matrix basis $\mathfrak{h}_n$
we obtain the matrix elements of the $L_L$ superoperator
\begin{equation}
\left(L_L\right)_{i,j} 
= \frac{1}{2}\sum_{k,l}\Gamma_{k,l}
\left(
2\tr\left[h_ih_l h_j h_k\right]
-\tr\left[h_i\{h_k h_l,h_j\}\right]
\right),
\label{eq:liouvilleLpart03}
\end{equation}
To explicitly calculate the first term of the previous equation
we therefore need to put traces of the form
$\tr[h_ih_l h_j h_k]$ in terms of the structure constants of
the Lie and Jordan algebras. Using the Hilbert-Schmidt inner product,
the properties of the trace
and the fact that any product of the form $h_k h_i$ can itself be expanded
in terms of $\mathfrak{h}_n$ as
\begin{equation}
h_k h_i =\sum_{p}\tr\left[h_k h_i h_p\right]h_p,
\end{equation}
we get
\begin{equation}
\tr\left[h_ih_l h_j h_k\right]
=\tr\left[h_k h_i h_l h_j \right]
=\sum_{p}\tr\left[h_k h_i h_p\right]\sum_{q}\tr\left[h_l h_j h_q\right]\tr\left[h_ph_q\right]
=\sum_{p}\tr\left[h_k h_i h_p\right]\tr\left[h_l h_j h_p\right].
\label{eq:trfourh00}
\end{equation}
The structure constants enter this expression through
\begin{equation}
\tr\left[h_k h_i h_p\right]=\frac{1}{2}\tr\left[
\left\{h_k, h_i\right\} h_p
+\left[h_k, h_i\right] h_p
\right]
=\frac{1}{2}\left(
b_{k,i,p}+ic_{k,i,p}
\right).\label{eq:tr3h00}
\end{equation}
Substituting the previous result into Eq.~\eqref{eq:trfourh00},
we obtain
\begin{equation}
\tr\left[h_i h_l h_j h_k\right]
=\frac{1}{4}\sum_p
\left(b_{k,i,p}+ic_{k,i,p}\right)
\left(b_{l,p,j}-ic_{l,p,j}\right)
=\frac{1}{4}\left[\left(B_k+iC_k\right)\left(B_l-iC_l\right)\right]_{i,j},
\label{eq:trfourh01}
\end{equation}
where, in the final expression, we have used Eq.~\eqref{eq:struconsc00}
and have defined the superoperators corresponding to the Jordan
structure constants
\begin{equation}
\left(B_i\right)_{j,k}=b_{i,j,k}.
\end{equation}
To simplify subsequent expressions and unify the Jordan and Lie
structure constants into a single object, it is convenient to define
\begin{equation}
z_{i,j,k} = 2\tr\left[h_i h_j h_k\right]
= b_{i,j,k}+ic_{i,j,k},
\label{eq:zdef00}
\end{equation}
and the corresponding Hermitian structure-constant superoperator
\begin{equation}
Z_k = B_k + iC_k,
\label{eq:zdef}
\end{equation}
which is Hermitian because \(B_k\) is real and symmetric,
whereas \(C_k\) is real and antisymmetric, as discussed above.
These superoperators play a central role in the algebraic formulation
developed throughout this work, as they serve as the fundamental
building blocks for the construction of the Liouville superoperator.
Furthermore, primarily from a computational standpoint,
it is convenient to define the third-rank tensor
\begin{equation}
\boldsymbol{Z}
=
\left(Z_1, Z_2, \dots, Z_n\right)^\top.
\end{equation}
This notation simplifies Eq. \eqref{eq:trfourh01}
to
\begin{equation}
\tr\left[h_i h_l h_j h_k\right]
=\frac{1}{4}\left(Z_kZ_l^*\right)_{i,j},
\label{eq:trfourh02}
\end{equation}

To evaluate the second term of $L_L$ in Eq. \eqref{eq:liouvilleLpart03},
we could, in principle, use Eq. \eqref{eq:trfourh02}. However, it is
more illuminating to compute it directly:
\begin{multline}
\tr\left[h_i\left\{h_kh_l,h_j\right\}\right]
=\frac{1}{2}\tr\left[h_i\left\{
\left\{h_k,h_l\right\}+\left[h_k,h_l\right],
h_j\right\}\right]
=\frac{1}{2}\tr\left[h_i\left\{
\sum_p\left(b_{k,l,p}+ic_{k,l,p}\right)h_p,
h_j\right\}\right]\\
=\frac{1}{2}\sum_p\left(b_{k,l,p}+ic_{k,l,p}\right)
\tr\left[h_i\left\{h_p,h_j\right\}\right]
=\frac{1}{2}\sum_p z_{k,l,p}b_{i,p,j}
=\frac{1}{2}\sum_p z_{k,l,p}b_{p,i,j}
=\frac{1}{2}\sum_p z_{k,l,p}\left(B_p\right)_{i,j}.
\label{eq:trfourh03}
\end{multline}

Using this result together with
Eqs. \eqref{eq:trfourh00} and \eqref{eq:trfourh02},
we can rewrite Eq. \eqref{eq:liouvilleLpart03} as
\begin{equation}
L_L = \frac{1}{4}\sum_{k,l}\Gamma_{k,l}
\left(Z_kZ_l^*
-\sum_p z_{k,l,p}B_p\right).
\end{equation}

Combining the Hamiltonian and Lindbladian contributions to
the Liouville superoperator, we obtain
\begin{equation}
L=\sum_{p}\left(
\frac{H_p}{\hbar}C_p -\frac{1}{4}\sum_{k,l}\Gamma_{k,l}z_{k,l,p}B_p
\right)
+ \frac{1}{4}\sum_{k,l}\Gamma_{k,l}Z_k Z_l^*.
\label{eq:liouvillefull00}
\end{equation}
In the following section, we show that the superoperators
arising from the products of the structure constants, $Z_kZ_l^*$,
not only possess their own algebraic structure, but also form an inner product space
under the Hilbert--Schmidt inner product.
Given the central role these operators play in the algebraic
structure of the Lindblad equation, it is therefore convenient to define
\begin{equation}
X_{k,l} = \frac{1}{4}Z_k Z_l^*,
\label{eq:xdef}
\end{equation}
and express the entire Liouville superoperator in terms of these elements,
thereby representing it as a linear combination of elements
of a vector space with an orthogonal basis.

To do so, we first express $C_p$ and $B_p$ in terms of the operators $Z_kZ_l^*$.
Choosing the first element of $\mathfrak{h}_k$ to be proportional to the identity operator,
i.e., $h_1 = I/\sqrt{m}$, yields
$C_1=0$, $B_1=2I/\sqrt{m}$, and consequently
$Z_1=Z_1^* = 2I/\sqrt{m}$.
It then follows that
\begin{equation}
Z_k=\frac{\sqrt{m}}{2}Z_k Z_1^*=\frac{\sqrt{m}}{2}X_{k,1},
\end{equation}
and similarly,
\begin{equation}
Z_l^*=\frac{\sqrt{m}}{2}Z_1 Z_l^*=\frac{\sqrt{m}}{2}X_{1,l}.
\end{equation}

Substituting these relations into Eq.~\eqref{eq:liouvillefull00}, we obtain
\begin{multline}
L=\sum_{p}\left(
\frac{H_p}{\hbar}C_p -\frac{1}{4}\sum_{k,l}\Gamma_{k,l}z_{k,l,p}B_p
\right)
+ \frac{1}{4}\sum_{k,l}\Gamma_{k,l}Z_k Z_l^*\\
=\frac{i\sqrt{m}}{\hbar}\sum_{p}
H_p\left(X_{1,p}-X_{p,1}\right)
-\frac{\sqrt{m}}{4}
\sum_{p}\sum_{k,l}\Gamma_{k,l}z_{k,l,p}\left(X_{1,p}+X_{p,1}\right)
+ \sum_{k,l}\Gamma_{k,l}X_{k,l}.
\label{eq:liouville02}
\end{multline}
By grouping terms, the Liouville superoperator
can finally be expressed as
\begin{equation}
L=\sum_{k,l}\Lambda_{k,l}X_{k,l},
\label{eq:liouvillefull01}
\end{equation}
where $\Lambda_{k,l}$ are the elements of the
second-rank tensor
\begin{eqnarray}
\boldsymbol{\Lambda}
= \boldsymbol{\Gamma}
+\sqrt{m}\left[
\boldsymbol{e}\otimes
\left(i \boldsymbol{H}
-\frac{1}{4}\tr\left[\boldsymbol{\Gamma}\boldsymbol{Z}\right]\right)
-\left(i \boldsymbol{H}
+\frac{1}{4}\tr\left[\boldsymbol{\Gamma}\boldsymbol{Z}\right]\right)
\otimes \boldsymbol{e}
\right].
\label{eq:Lambdadef00}
\end{eqnarray}
Here, $\boldsymbol{e}=(1,0,\dots,0)^\top$
is an $n$-dimensional vector, $\boldsymbol{H}$
is the Hamiltonian vector defined in
Eq.~\eqref{eq:hamiltonianvector00},
and $\boldsymbol{\Gamma}$ is the tensor whose elements are
$\Gamma_{k,l}$.
The trace in the vector
$\tr\left[\boldsymbol{\Gamma}\boldsymbol{Z}\right]$
contracts only the first index of $\boldsymbol{\Gamma}$
with the second index of $\boldsymbol{Z}$,
thereby yielding a first-rank tensor.
From a computational perspective, defining the tensor
$\boldsymbol{\Lambda}$ is particularly useful. First,
its elements can be computed efficiently from the Hamiltonian
and dissipative coefficients. Second, all model-dependent
quantities are contained in $\boldsymbol{\Lambda}$,
whereas the superoperators $X_{k,l}$ depend only on the chosen basis.
Consequently, the latter need to be computed only once and can be
reused for different Hamiltonians and dissipative processes,
substantially reducing the computational cost of constructing the
Liouville superoperator.

Up to this point, we have succeeded in expressing the Liouville superoperator
associated with the Lindblad equation as
a linear combination of the basis elements $X_{k,l}$.
However, the algebraic structure of these elements,
as well as their orthogonality properties,
remain to be established.
This will be addressed in the next section.

\section{Algebraic structure}\label{sec:algebraicstructure}
The ultimate objective of this section is to uncover the algebraic
structure of the Lindblad equation.
This entails establishing the structure constants of the
algebra generated by the basis elements $X_{i,j}$.
Additionally, we show that, under certain conditions,
the set $\{X_{i,j}\}$ forms an orthonormal basis with respect
to the Hilbert--Schmidt inner product.

We begin by determining the algebra satisfied by the set of
structure constants $\{C_i\}$.
Although this task is typically carried out using the
Jacobi identity \cite{georgi2000lie},
we provide here an alternative demonstration that does
not rely on it.
This approach will be particularly useful for determining
the algebraic structure of the larger set $\{C_i,B_j\}$.

The algebra of the set \(\{C_i\}\) can be obtained
from the commutator
\begin{equation}
\left[C_i,C_j\right]_{\alpha,\beta}=
\sum_{k}\left(
c_{i,\alpha,k}c_{j,k,\beta} 
- c_{j,\alpha,k}c_{i,k,\beta}
\right)
=\sum_{k,l}\left(
c_{i,\alpha,k}c_{j,l,\beta}
- c_{j,\alpha,k}c_{i,l,\beta}
\right)\tr\left[h_k h_l\right],
\label{eq:algcc00}
\end{equation}
where we have used the orthonormality condition given in
Eq.~\eqref{eq:orthonormh00}.
With the aid of
Eqs.~\eqref{eq:struconst00} and \eqref{eq:struconst01},
the first and second terms in the previous equation
can be rewritten as
\begin{equation}
\sum_{k,l}
c_{i,\alpha,k}c_{j,l,\beta}
\tr\left[h_k h_l\right]
=-\tr\left[
\sum_k c_{i,\alpha,k}h_k
\sum_l
c_{j,\beta,l}h_l
\right]
=-\tr\left[\left[h_i,h_\alpha\right]\left[h_j,h_\beta\right]\right]
\end{equation}
and
\begin{equation}
\sum_{k,l}
c_{j,\alpha,k}c_{i,l,\beta}
\tr\left[h_k h_l\right]
=-\tr\left[
\sum_k c_{j,\alpha,k}h_k
\sum_l
c_{i,\beta,l}h_l
\right]
=-\tr\left[\left[h_j,h_\alpha\right]\left[h_i,h_\beta\right]\right],
\end{equation}
respectively.
Substituting these two results into Eq.~\eqref{eq:algcc00},
applying the properties of the trace, and collecting terms, we obtain
\begin{multline}
\left[C_i,C_j\right]_{\alpha,\beta}
=\tr\big[
h_i h_\alpha h_\beta h_j + h_\alpha h_i h_j h_\beta
-h_j h_\alpha h_\beta h_i - h_\alpha h_j h_i h_\beta
\big]
=-\tr\Big[
\left[h_i, h_j\right]\left[h_\alpha h_\beta\right]
\Big]\\
=-\tr\Big[
\sum_k \tr\Big[\left[h_i, h_j\right]h_k\Big]h_k
\sum_l \tr\Big[h_\alpha h_\beta h_l\Big] h_l
\Big]
=-\sum_k c_{i,j,k}c_{\alpha,\beta,k}
=-\sum_k c_{i,j,k}c_{k,\alpha,\beta}.
\end{multline}
Summarizing the previous result,
\begin{equation}
\left[C_i,C_j\right]=-\sum_k c_{i,j,k}C_k,
\label{eq:algcc01}
\end{equation}
thus, the structure constants themselves furnish a representation
of the algebra of \(\mathfrak{h}_n\), with the nuance that,
compared with Eq.~\eqref{eq:struconst00}, here \(i c_{i,j,k}\)
is replaced by \(-c_{i,j,k}\).

Following a similar procedure, we can further calculate the
commutator for the Jordan structure constants,
\begin{equation}
\left[B_i,B_j\right] = \sum_{k}c_{i,j,k}C_k.
\label{eq:algbb01}
\end{equation}
Therefore, the set \( \{B_j\} \) does not form an algebra by itself,
since the commutator of its elements lies in the set \( \{C_i\} \).
However, by computing
the commutators of the crossed terms between
the Lie \( \{C_i\} \) and Jordan \( \{B_j\} \) structure
constants, we obtain
\begin{eqnarray}
\left[C_i,B_j\right] &=& -\sum_{k}c_{i,j,k}B_k,\label{eq:algcb01}\\
\left[B_i,C_j\right] &=& -\sum_{k}c_{i,j,k}B_k.\label{eq:algbc01}
\end{eqnarray}
This means that \( \{C_i, B_j\} \) does in fact form an algebra,
although it is not a realization of the algebra of \(\mathfrak{h}_n\).

It is natural to compute the commutators arising from the sets
\( \{Z_i\} \) and \( \{Z_j^*\} \), defined in
Eq.~\eqref{eq:zdef}, since these superoperators provide
a convenient representation of the Liouville superoperator,
as can be seen from Eqs.~\eqref{eq:liouvillefull00},
\eqref{eq:xdef}, and \eqref{eq:liouvillefull01}.
Using Eqs.~\eqref{eq:zdef} and \eqref{eq:algcc01}--\eqref{eq:algbc01},
we obtain the following commutators:
\begin{eqnarray}
\left[Z_i, Z_j\right] &=& -2i\sum_k c_{i,j,k}Z_k,\\
\left[Z_i^*, Z_j^*\right] &=& 2i\sum_k c_{i,j,k}Z_k^*,\\
\left[Z_i, Z_j^*\right] &=& 0.
\end{eqnarray}
The sets \(\mathfrak{Z}_n = \{Z_i\} \) and
\( \mathfrak{Z}_n^* = \{Z_j^*\} \) therefore
constitute two independent realizations
of the Lie algebra associated with \(\mathfrak{h}_n\).
With these results at hand, we can finally determine the
algebra of the set $\mathfrak{X}_{n^2}=\{X_{i,j}\}$, which completely
determines the Liouville operator in Eq.~\eqref{eq:liouvillefull01}.
This requires calculating the commutator
$[X_{i,j}, X_{i^\prime,j^\prime}]$ and the anticommutator
$\{X_{i,j}, X_{i^\prime,j^\prime}\}$,
and expressing them as linear combinations
of the elements of $\mathfrak{X}_{n^2}$.
To this end, we use the definition of
$X_{i,j}$ in Eq.~\eqref{eq:xdef}
together with Eq.~\eqref{eq:trfourh02},
\begin{equation}
\left[X_{i,j}, X_{i^\prime,j^\prime}\right]_{\alpha,\beta}
=\sum_k\left(
\tr\left[h_i h_\alpha h_j h_k\right]
\tr\left[h_{i^\prime} h_k  h_{j^\prime} h_\beta\right]
-\tr\left[h_{i^\prime} h_\alpha h_{j^\prime} h_k\right]
\tr\left[h_i h_k h_j h_\beta\right]
\right).
\end{equation}
he elements appearing in the products of traces on the right-hand side
of the previous equation can be combined
by expanding products of $\mathfrak{h}_n$ elements and using
$\delta_{k,l}=\tr[h_k h_l]$,
\begin{multline}
\left[X_{i,j}, X_{i^\prime,j^\prime}\right]_{\alpha,\beta}
=\sum_k\left(
\tr\left[h_i h_\alpha h_j h_k\right]
\tr\left[ h_{j^\prime} h_\beta h_{i^\prime}h_k\right]
-\tr\left[h_{i^\prime} h_\alpha h_{j^\prime} h_k\right]
\tr\left[ h_j h_\beta h_i h_k\right]
\right)\\
=  \tr\left[
h_i h_\alpha h_j h_{j^\prime} h_\beta h_{i^\prime}
-h_{i^\prime} h_\alpha h_{j^\prime} h_j h_\beta h_i
\right]
= \tr\left[
h_{i^\prime} h_i h_\alpha h_j h_{j^\prime} h_\beta
-h_i h_{i^\prime} h_\alpha h_{j^\prime} h_j h_\beta
\right].
\end{multline}
This form is particularly convenient because the traces can
be recursively decomposed into lower-order traces through
expansions in the basis elements, making the underlying
structure constants explicit. One of the many possible
ways of factorizing the traces is
\begin{multline}
\left[X_{i,j}, X_{i^\prime,j^\prime}\right]_{\alpha,\beta}
= \tr\left[
\left(\sum_k \tr\left[h_{i^\prime} h_i h_k\right]h_k\right)
h_\alpha h_j h_{j^\prime} h_\beta
-\left(\sum_k \tr\left[h_i h_{i^\prime}h_k\right]h_k\right)
h_\alpha h_{j^\prime} h_j h_\beta
\right]\\
=\sum_k\left( \tr\left[h_{i^\prime} h_i h_k\right]
\tr\left[h_k
h_\alpha h_j h_{j^\prime} h_\beta
\right]
-\tr\left[h_i h_{i^\prime}h_k\right]
\tr\left[h_k
h_\alpha h_{j^\prime} h_j h_\beta
\right]
\right).
\label{eq:commuX00}
\end{multline}
We can readily identify the triple traces of the form
\begin{equation}
\tr\left[h_i h_{i^\prime} h_k\right]
=-\tr\left[h_{i^\prime} h_i h_k\right]
=\frac{1}{2}z_{i,i^\prime,k}
=\frac{1}{2}\left(b_{i,i^\prime,k}+ic_{i,i^\prime,k}\right)
=\frac{1}{2}\left(Z_k\right)_{i,i^\prime},
\label{eq:zdef00}
\end{equation}
using Eq. \eqref{eq:tr3h00}.
The quintuple traces can be further decomposed as
\begin{multline}
\tr\left[
h_k h_\alpha h_j h_{j^\prime} h_\beta
\right]
=\tr\left[
h_j h_{j^\prime} h_\beta h_k h_\alpha
\right]
=\tr\left[
\left(\sum_l \tr\left[h_j h_{j^\prime}h_l\right]\right)h_l
 h_\beta h_k h_\alpha
\right]\\
=\frac{1}{2}\sum_l z_{j,j^\prime, l}
\tr\left[h_l
h_\beta h_k h_\alpha\right]
=\frac{1}{2}\sum_l z_{j,j^\prime, l}
\tr\left[ h_k h_\alpha h_l
 h_\beta\right]
 =\frac{1}{2}\sum_l z_{j,j^\prime, l}
 \left(X_{k,l}\right)_{\alpha,\beta}.
\end{multline}
The structure constants of the Lie algebra
generated by $\mathfrak{X}_{n^2}$ are obtained by
substituting this result together with Eq.~\eqref{eq:zdef00}
into Eq.~\eqref{eq:commuX00}, yielding
\begin{equation}
\left[X_{i,j}, X_{i^\prime,j^\prime}\right]_{\alpha,\beta}
=\frac{1}{4}\sum_{k,l}\left(
z_{i,i^\prime,k}^* z_{j,j^\prime,l}
-z_{i,i^\prime,k} z_{j,j^\prime,l}^*
\right)\left(X_{k,l}\right)_{\alpha,\beta},
\end{equation}
or, more succinctly,
\begin{equation}
\left[X_{i,j}, X_{i^\prime,j^\prime}\right]
=\frac{1}{4}\sum_{k,l}\left(
z_{i,i^\prime,k}^* z_{j,j^\prime,l}
-z_{i,i^\prime,k} z_{j,j^\prime,l}^*
\right)X_{k,l}.
\label{eq:liextwoindex00}
\end{equation}
Similarly, the Jordan algebra structure constants
that arise from $\mathfrak{X}_{n^2}$ come about
from the anticommutator. A calculation similar
to the previous one yields
\begin{equation}
\left\{X_{i,j}, X_{i^\prime,j^\prime}\right\}
=\frac{1}{4}\sum_{k,l}\left(
z_{i,i^\prime,k}^* z_{j,j^\prime,l}
+z_{i,i^\prime,k} z_{j,j^\prime,l}^*
\right)X_{k,l}.
\label{eq:jordanxtwoindex00}
\end{equation}

Equations \eqref{eq:liextwoindex00} and \eqref{eq:jordanxtwoindex00}
show that the set $\mathfrak{X}_{n^2}$
is closed under both the commutator and anticommutator operations.
In particular, closure under the commutator implies that
$\mathfrak{X}_{n^2}$ forms a Lie algebra.
Consequently, the evolution operator associated with the
Lindblad equation \eqref{eq:liouvillefull01} belongs to the
corresponding Lie group and may be parametrized using
either the canonical coordinates of the first kind
\cite{wei1963,varadarajan2013lie,sandoval2019},
\begin{equation}
V(t)=\prod_{i,j} \exp\left[\parprod_{i,j}(t)X_{i,j}\right],
\label{eq:dynamicalmap01}
\end{equation}
or the canonical coordinates of the second kind,
\begin{equation}
V(t)=\exp\left[\sum_{i,j} \parsum_{i,j}(t)X_{i,j}\right],
\label{eq:dynamicalmap02}
\end{equation}
where the functions $\parprod_{i,j}(t)$ and
$\parsum_{i,j}(t)$ remain
to be determined.
Methods have been developed to derive the differential
equations associated with each of these representations
from the Liouville equation \eqref{eq:Lindblad01}
\cite{wei1963,sandoval2019}.
Moreover, techniques other than the
Baker-Campbell-Hausdorff formulas have been developed
to establish a correspondence between the parameters
$\parprod_{i,j}(t)$ and $\parsum_{i,j}(t)$ \cite{sandoval2019}.
However, these approaches generally lead to nonlinear
differential equations whose complexity grows rapidly
with the dimension of the system's Hilbert space.

It is important to note that, in the special case of the
von Neumann equation, when $\Gamma=0$, the first line of
Eq.~\eqref{eq:liouville02} shows that the only generators
of the group that contribute are the structure constants $C_i$,
which form a subalgebra of $X_{i,j}$.
Therefore, in this case, the dynamical map can
be parametrized in terms of only $n-1$
(recall that the structure constants corresponding to
$h_1=I/2^{n_q/2}$ satisfy $C_1=0$.)
Therefore, in this case, the dynamical map can be parametrized in terms of only
$n-1$ functions as
\begin{equation}
V(t)=\prod_{i}\exp\left[\parprod_{i}(t)C_{i}\right],
\end{equation}
or, alternatively,
\begin{equation}
V(t)=\exp\left[\sum_{i}\parsum_{i}(t)C_{i}\right].
\end{equation}
This reveals that the Lindbladian increases the complexity
of the Liouville equation by requiring a larger set of generators.

In the next section, we show that, under certain conditions,
the set $\mathfrak{X}_{n^2}$ forms an inner product space
whose elements are orthogonal with respect to the
Hilbert-Schmidt inner product.
This property greatly simplifies the parametrization
of the evolution operator.

\section{The inner-product space $\mathfrak{X}_{n^2}$}
\label{sec:innerproductspaceX}
So far, the only restrictions imposed on the matrix
basis $\mathfrak{h}_n$ are that it be composed of
Hermitian matrices that are orthonormal with respect
to the Hilbert-Schmidt inner product.
A wide variety of sets can therefore be used as
$\mathfrak{h}_n$.
A convenient way to construct such a basis is to
build it recursively from the set consisting of
the normalized identity matrix and the Pauli matrices.
The Pauli matrices and their tensor products,
commonly known as Pauli strings,
play a central role in quantum physics and quantum
information theory.
Together with the identity matrix, they form a complete
orthonormal basis for the space of operators acting on
qubit systems, making them particularly suitable for the
representation of quantum states, observables, Hamiltonians,
and quantum channels.
Their importance extends to quantum simulation,
quantum error correction, stabilizer codes,
and quantum algorithms, where the decomposition of
operators into Pauli strings often constitutes a
fundamental step in both analytical and numerical
approaches
\cite{gottesman1998, nielsen2010quantum, georgescu2014,
mcardle2020, Kunold_2024, Hantzko_2024}.
Given the importance of Pauli strings, considerable efforts
have been devoted to optimizing the Pauli decomposition
process \cite{Hantzko_2024}, which is essential for determining
the coefficients appearing in Eq.~\eqref{eq:liouvillefull01}
and often entails a substantial computational cost.

In this section, we focus on the properties of Pauli strings.
It should be emphasized, however, that the results derived
below are not exclusive to Pauli strings and apply more
generally to other matrix basis satisfying the assumptions
introduced previously.

For a system of $n_q$ qubits, the corresponding Pauli strings
are constructed as tensor products of Pauli matrices and
the identity operator,
\begin{equation}
h_i^{n_q}=\frac{1}{2^{n_q/2}}
\sigma_{i_{n_q}}\otimes
\sigma_{i_{n_q-1}}\otimes\cdots\otimes
\sigma_{i_1},
\end{equation}
where $\sigma_1$, $\sigma_2$, and $\sigma_3$ denote
the Pauli matrices, $\sigma_0$ is the identity matrix,
and the factor $2^{-n_q/2}$ ensures Hilbert-Schmidt
normalization. The index $i$ can be conveniently identified with the
integer whose binary representation is given by the string
$i_{n_q}i_{n_q-1}\cdots i_1$. Explicitly,
\begin{equation}
i=1+\sum_{n=1}^{n_q}2^{\,n-1}i_n,
\end{equation}
where $i_n\in\{0,1\}$, so that
$i=1,2,\ldots,2^{n_q}$.
Following this notation, the elements $h_i^1$
are simply the normalized identity matrix and
the normalized Pauli matrices.
This construction provides a recursive procedure
for generating the elements of an $n_q$-qubit basis:
\begin{equation}
h_i^{n_q+1}
=h_{i_2}^{n_q}\otimes h_{i_1}^{1},
\label{eq:recforh00}
\end{equation}
where we have introduced a superscript on the
elements of $\mathfrak{h}_n$, with $n=2^{n_q}$,
to indicate the number of qubits on which they act.
Similarly, one can derive recursion relations for the Lie and Jordan
structure constants \cite{Kunold_2024}.
For our purposes, the recursion relations for
$Z_k$ and $X_{i,j}$ are of particular interest.
Substituting the recursion relation in Eq.~\eqref{eq:recforh00}
into the definition of $z_{i,j,k}$ given in Eq.~\eqref{eq:zdef00},
we obtain
\begin{equation}
z_{i,j,k}^{n_q+1}
=2 \tr\left[
h_{i_2}^{n_q}h_{j_2}^{n_q}h_{k_2}^{n_q}
\otimes h_{i_1}^{1}h_{j_1}^{1}h_{k_1}^{1}
\right]
=2 \tr\left[
h_{i_2}^{n_q}h_{j_2}^{n_q}h_{k_2}^{n_q}
\right]
\tr\left[
h_{i_1}^{1}h_{j_1}^{1}h_{k_1}^{1}
\right]
=\frac{1}{2}
z_{i_2,j_2,k_2}^{n_q}
z_{i_1,j_1,k_1}^{1},
\end{equation}
and, after flattening the indices, we obtain
\begin{equation}
Z_{k}^{n_q+1}
=\frac{1}{2}Z_{k_2}^{n_q}\otimes Z_{k_1}^{1},
\end{equation}
or, in tensorial notation
\begin{equation}
\boldsymbol{Z}^{n_q+1}
=\frac{1}{2}\boldsymbol{Z}^{n_q}\otimes \boldsymbol{Z}^{1}.
\end{equation}
The structure constants $B_i$ and $C_i$ satisfy similar recursion
relations,
\begin{eqnarray}
C_k^{n_q+1} = \frac{1}{2}\left(C_{k_2}^{n_q}\otimes B_{k_1}^1
+ B_{k_2}^{n_q}\otimes C_{k_1}^1\right),\\
B_k^{n_q+1} = \frac{1}{2}\left(B_{k_2}^{n_q}\otimes B_{k_1}^1
- C_{k_2}^{n_q}\otimes C_{k_1}^1\right),
\end{eqnarray}
or, in tensorial notation,
\begin{eqnarray}
\boldsymbol{C}^{n_q+1}
= \frac{1}{2}\left(\boldsymbol{C}^{n_q}\otimes \boldsymbol{B}^1
+ \boldsymbol{B}^{n_q}\otimes \boldsymbol{C}^1\right),\\
\boldsymbol{B}^{n_q+1}
= \frac{1}{2}\left(\boldsymbol{B}^{n_q}\otimes \boldsymbol{B}^1
- \boldsymbol{C}^{n_q}\otimes \boldsymbol{C}^1\right).
\end{eqnarray}
Although these recursion relations are not used explicitly in the
present work, they are of independent interest.
The accompanying Mathematica notebook includes functions
for computing these structure constants recursively.
The derivation of these expressions can be found
in Appendix \ref{app:a}

Using this recursion relation together with the definition of
$X_{i,j}$ in Eq.~\eqref{eq:xdef},
we find
\begin{equation}
X_{i,j}^{n_q+1}
=\frac{1}{4}Z_i^{n_q+1}Z_j^{n_q+1 *}
=\frac{1}{16}Z_{i_2}^{n_q}\otimes Z_{i_1}^{1}
Z_{j_2}^{n_q *}\otimes Z_{j_1}^{1 *}
=\frac{1}{4}Z_{i_2}^{n_q} Z_{j_2}^{n_q *}
\otimes \frac{1}{4}Z_{i_1}^{1}Z_{j_1}^{1 *}
=X_{i_2,j_2}^{n_q}
\otimes X_{i_1,j_1}^{1}
\end{equation}
which can be written compactly
in tensor-product notation
as
\begin{equation}
\boldsymbol{X}^{n_q+1}
=\boldsymbol{X}^{n_q}
\otimes \boldsymbol{X}^{1}.
\end{equation}
These recursion relations are particularly useful from a
computational perspective, as they allow the structure
constants of an $n_q$-qubit basis to be constructed
iteratively from those of lower-dimensional bases,
thereby avoiding the direct evaluation of large tensor
products and traces.
Beyond their computational utility, these recursion relations
also provide a convenient framework for establishing
the orthonormality of $\mathfrak{X}_{n^2}$ by induction.

To this end, we consider the Hilbert-Schmidt inner product
of two elements of $\mathfrak{X}_{n^2}$.
Using the recursion relation for
$X_{i,j}^{n_q+1}$, we obtain
\begin{equation}
\tr\left[X_{i,j}^{n_q+1}X_{i^\prime,j^\prime}^{n_q+1}\right]
=\tr\left[X_{i_2,j_2}^{n_q}X_{i_2^\prime,j_2^\prime}^{n_q}\right]
\tr\left[X_{i_1,j_1}^{1}X_{i_1^\prime,j_1^\prime}^{1}\right].
\label{eq:orthoXX00}
\end{equation}
By direct evaluation of all $256$ possible index combinations,
one finds that
\begin{equation}
\tr\left[X_{i_1,j_1}^{1}X_{i_1^\prime,j_1^\prime}^{1}\right]
=\delta_{i_1,i_1^\prime}\delta_{j_1,j_1^\prime}.
\end{equation}
Consequently, for the two-qubit case,
\begin{equation}
\tr\left[X_{i,j}^{2}X_{i^\prime,j^\prime}^{2}\right]
=\tr\left[X_{i_2,j_2}^{1}X_{i_2^\prime,j_2^\prime}^{1}\right]
\tr\left[X_{i_1,j_1}^{1}X_{i_1^\prime,j_1^\prime}^{1}\right]
=\delta_{i_2,i_2^\prime}\delta_{j_2,j_2^\prime}
\delta_{i_1,i_1^\prime}\delta_{j_1,j_1^\prime}
=\delta_{i,i^\prime}\delta_{j,j^\prime}.
\end{equation}
Assuming that
\begin{equation}
\tr\left[X_{i,j}^{n_q}X_{i^\prime,j^\prime}^{n_q}\right]
=\delta_{i,i^\prime}\delta_{j,j^\prime},
\end{equation}
Eq.~\eqref{eq:orthoXX00} immediately implies that
\begin{equation}
\tr\left[X_{i,j}^{n_q+1}X_{i^\prime,j^\prime}^{n_q+1}\right]
=\delta_{i,i^\prime}\delta_{j,j^\prime}.
\end{equation}
Therefore, by mathematical induction,
\begin{equation}
\tr\left[X_{i,j}^{n_q}X_{i^\prime,j^\prime}^{n_q}\right]
=\delta_{i,i^\prime}\delta_{j,j^\prime}
\end{equation}
for all $n_q$.
Since \(\mathfrak{X}_{n^2}\) contains \(n^2\)
mutually orthonormal Hermitian operators and spans the
\(n^2\)-dimensional space of all \(n\times n\) matrices,
it forms an orthonormal Hermitian basis for this space.

In the next section, we combine the algebraic, vector-space,
and inner-product structures of \(\mathfrak{X}_{n^2}\)
to derive explicit expressions for the Liouville superoperator
and the associated dynamical map \(V(t)\).

\section{The dynamical map}\label{sec:dynamicalmap}
Since $\mathfrak{X}_{n^2}$ forms an orthonormal Hermitian basis,
the dynamical map can be expanded as
\begin{equation}
V(t)=\sum_{i,j}V_{i,j}(t)X_{i,j},
\end{equation}
where
\begin{equation}
V_{i,j}(t)=\tr\left[V(t)X_{i,j}\right].
\end{equation}
The coefficients $V_{i,j}(t)$ are the expansion coefficients,
or equivalently the coordinates, of the superoperator
$V(t)$ in the basis $\mathfrak{X}_{n^2}$.
Collecting these coefficients into a vector,
\begin{equation}
\boldsymbol{V}(t)=
\left(
V_{1,1}(t),V_{1,2}(t),\ldots,
V_{2,1}(t),V_{2,2}(t),\ldots,
V_{n,n}(t)
\right)^\top,
\end{equation}
they should not be interpreted as the matrix elements
of the superoperator,
$\tr\left[h_iV(t)h_j\right]$.
It is worth noting that this expansion,
which is characterized by the $n^2$ coefficients
$V_{i,j}(t)$, is equivalent to those in
Eqs.~\eqref{eq:dynamicalmap01} and
\eqref{eq:dynamicalmap02}, which are
parameterized by the $n^2$ coefficients
$\alpha_{i,j}(t)$ and $\beta_{i,j}(t)$,
respectively.
Just as there exist maps relating the
parameters $\alpha_{i,j}(t)$ and $\beta_{i,j}(t)$
\cite{wei1963,varadarajan2013lie,sandoval2019},
there should also exist maps relating the coefficients
$V_{i,j}(t)$ to each of these parameterizations.
The explicit construction of these maps lies beyond
the scope of the present work.

Substituting the definition of the dynamical map
given in Eq.~\eqref{eq:dynamicalmap00}
into Eq.~\eqref{eq:Lindblad01},
we find that, in close analogy with
closed quantum systems, where the
evolution operator obeys the
Schr\"odinger equation, the dynamical map
obeys the equation \cite{breuer321}
\begin{equation}
\frac{d}{dt}V(t)=LV(t).
\end{equation}

Inserting the expansion of the Liouville superoperator
in terms of the basis elements of
$\mathfrak{X}_{n^2}$ [Eq.~\eqref{eq:liouvillefull01}]
into the previous equation, together with the
expansion of the dynamical map, the Lindblad equation
takes the form
\begin{multline}
\sum_{i,j}\frac{d}{dt}V_{i,j}(t)X_{i,j}
=\sum_{k,l}\Lambda_{k,l}X_{k,l}\sum_{p,q}V_{p,q}(t)X_{p,q}
=\sum_{k,l}\sum_{p,q}\Lambda_{k,l}V_{p,q}(t)X_{k,l}X_{p,q}\\
=\frac{1}{2}\sum_{k,l}\sum_{p,q}\Lambda_{k,l}V_{p,q}(t)
\left(
\left[X_{k,l},X_{p,q}\right]+\left\{X_{k,l},X_{p,q}\right\}
\right).
\end{multline}
The last line illustrates the usefulness of the
Lie and Jordan algebraic structures previously derived
for $\mathfrak{X}_{n^2}$ in
Eqs.~\eqref{eq:liextwoindex00} and
\eqref{eq:jordanxtwoindex00}.
Indeed, substituting both of these
relations into the previous equation,
the Lindblad equation can be readily
expanded in the basis $\mathfrak{X}_{n^2}$ as
\begin{equation}
\sum_{i,j}\frac{d}{dt}V_{i,j}(t)X_{i,j}
=\frac{1}{4}\sum_{k,l}\sum_{p,q}\sum_{r,s}
\Lambda_{k,l}V_{p,q}(t)
z_{k,p,r}^*z_{l,q,s}
X_{r,s},
\end{equation}
from which we obtain the evolution equation for the coordinate vector 
$\boldsymbol{V}(t)$
\begin{equation}
\frac{d}{dt}V_{i,j}(t)
=\frac{1}{4}\sum_{k,l}\sum_{p,q}
\Lambda_{k,l}V_{p,q}(t)
z_{k,p,i}^*z_{l,q,j}.
\end{equation}

The previous expression is not only a compact
and completely general representation
of the Lindblad equation, but also provides a significant
computational advantage over the direct approach.
The reason is that the algebraic formulation separates
the basis-dependent and model-dependent parts of the calculation.
The direct method requires explicit projections
of the Liouville map onto the matrix basis.
This involves repeated products and contractions
of high-rank tensors constructed from the basis elements.
In contrast, once the tensors $\boldsymbol{Z}$ and
$\boldsymbol{X}$ have been constructed,
the model-dependent information enters only through
$\boldsymbol{\Lambda}$, and the Liouville superoperator
is obtained by a single contraction.
This reduces the number of repeated tensor products
and makes the algebraic method faster, especially when
the same basis is reused for different Hamiltonians or
dissipative processes.

In the next section, we present the one-qubit case as an illustrative example.
We explicitly construct the matrix basis, compute the corresponding
structure-constant superoperators, and express the Liouville superoperator
in terms of them.
We then show that the Liouville superoperator and the differential equations
derived from the algebraic formulation are fully equivalent to those
obtained through the direct approach.

\section{Examples}\label{seq:examples}
A convenient system for illustrating the results developed in this work
is the transmon qubit. Since the number of quantum levels retained in the model
can be chosen arbitrarily, it provides a simple framework for studying the behavior
of the algebraic structures introduced above as the Hilbert-space dimension increases.
The corresponding Mathematica notebook can be downloaded from \cite{lindbladalgebra}.
The Hamiltonian is given by
\begin{equation}
H=\hbar\omega_0 a^\dagger a
+\frac{\hbar \alpha}{2}a^{\dagger 2} a^2
+\frac{\hbar \Omega}{2}\left(
a e^{i\omega t}
+a^\dagger e^{-i\omega t}
\right),
\end{equation}
where $a$ and $a^\dagger$ are the lowering and raising operators,
respectively, $\omega_0$ is the fundamental transition frequency,
$\alpha$ is the anharmonicity, $\Omega$ is the driving amplitude,
and $\omega$ is the driving frequency.
The first term describes a harmonic oscillator, the second introduces
the weak anharmonicity characteristic of the transmon, and the third
accounts for coherent driving by an external microwave field.
For the sake of simplicity, we assume that
the only allowed transitions are those
between neighbouring quantum levels,
described by the jump operators
$A_k = \left\vert k\right\rangle \left\langle k+1\right\vert$.
Additionally, we assume that all these transitions
are characterized by the same decay rate $\gamma$.
Under these assumptions, the Lindbladian is given by
\begin{equation}
L_L\rho_S(t)=
\frac{1}{2}\gamma\sum_{k}
\left[
2A_k\rho_S(t)A_k^\dagger
-\left\{A_k^\dagger A_k,\rho_S(t)\right\}
\right].
\end{equation}
To obtain solutions with reduced oscillatory behavior,
it is convenient to move to the rotating frame
through the transformation
\begin{equation}
U_0=\exp\left\{it\left[-\left(\omega_0+\frac{\alpha}{2}\right)a^\dagger a
+\frac{\alpha}{2}\left(a^\dagger a\right)^2\right]\right\}.
\end{equation}
In the rotating frame, the Lindblad equation takes the form
\begin{equation}
\frac{d}{dt}\rho(t) = -\frac{i}{\hbar}\left[H_R,\rho(t)\right]
+\frac{1}{2}\gamma\sum_{k}
\left[
2A_k\rho(t)A_k^\dagger
-\left\{A_k^\dagger A_k,\rho(t)\right\}
\right],
\end{equation}
where the Hamiltonian and density matrix in the rotating frame
are given by
\begin{equation}
H_R = \frac{\hbar \Omega}{2}\left(
a e^{i\Delta t}
+a^\dagger e^{-i\Delta t}
\right),
\end{equation}
and
\begin{equation}
\rho(t) = U_0\rho_S(t)U_0^\dagger,
\end{equation}
respectively. Here,
$\Delta=\omega-\omega_0+\alpha n $
denotes the detuning.

In the following, we consider the one-qubit transmon
as an example to illustrate the algebraic method for
deriving the dynamical equations corresponding to the
Lindblad equation.
The one-qubit system yields sufficiently
simple expressions that can be readily visualized.

Since the algebraic quantities depend only on the
underlying Hilbert space and not on the particular
Hamiltonian or dissipative processes, we begin by
constructing the one-qubit basis and its associated
structure-constant superoperators. These quantities
can subsequently be used to build the Liouville
superoperator of the transmon.
The required Hermitian basis is
\begin{align}
& h_1 = \frac{1}{\sqrt{2}}\left(
\begin{array}{cc}
1 & 0 \\
0 & 1 \\
\end{array}\right),
& h_2 = \frac{1}{\sqrt{2}}\left(
\begin{array}{cc}
0 & 1 \\
1 & 0 \\
\end{array}\right), & \\
&h_3 = \frac{1}{\sqrt{2}}\left(
\begin{array}{cc}
0 & -i \\
i & 0 \\
\end{array}\right),
&h_4 = \frac{1}{\sqrt{2}}\left(
\begin{array}{cc}
1 & 0 \\
0 & -1 \\
\end{array}\right). &
\end{align}

The corresponding Hermitian structure-constant
superoperators are
\begin{align}
& Z_1= \sqrt{2}\left(
\begin{array}{cccc}
1 & 0 & 0 & 0 \\
 0 & 1 & 0 & 0 \\
 0 & 0 & 1 & 0 \\
 0 & 0 & 0 & 1 \\
\end{array}
\right)\,\, ,
&Z_2=\sqrt{2}\left(
\begin{array}{cccc}
 0 & 1 & 0 & 0 \\
 1 & 0 & 0 & 0 \\
 0 & 0 & 0 & i \\
 0 & 0 & -i & 0 \\
\end{array}
\right)\,\, ,\\
&Z_3=\sqrt{2}\left(
\begin{array}{cccc}
 0 & 0 & 1 & 0 \\
 0 & 0 & 0 & -i \\
 1 & 0 & 0 & 0 \\
 0 & i & 0 & 0 \\
\end{array}
\right)\,\, ,
&Z_4=\sqrt{2}\left(
\begin{array}{cccc}
 0 & 0 & 0 & 1 \\
 0 & 0 & i & 0 \\
 0 & -i & 0 & 0 \\
 1 & 0 & 0 & 0 \\
\end{array}
\right)\,\, .
\end{align}
The Lie and Jordan structure constants can be
readily obtained from these matrices.
In this example, the set \(\mathfrak{X}_{n^2}\)
consists of 16 elements, which we do not list
here but which can be visualized in the
accompanying Mathematica notebook.

To obtain the Liouville superoperator
in the form of Eq.~\eqref{eq:liouvillefull01},
it is first necessary to compute the tensor
\(\Lambda\), which in turn requires evaluating
the Hamiltonian and Lindbladian contributions.

We first consider the Hamiltonian contribution to the
Liouville superoperator.
Truncating the transmon Hamiltonian
in the interaction picture to its first
two quantum levels yields
\begin{equation}
H_i = \frac{\Omega}{2}\left(
\begin{array}{cc}
0 & \mathrm{e}^{i \Delta t}\\
\mathrm{e}^{-i \Delta t} & 0 \\
\end{array}
\right).
\end{equation}
The corresponding Hamiltonian vector
is obtained by projecting onto the elements
of $\mathfrak{h}_n$,
\begin{equation}
\boldsymbol{H}=\left(0,\frac{\Omega}{2}\cos(\Delta t),
-\frac{\Omega}{2}\sin(\Delta t),0
\right)^\top
\end{equation}

We now turn to the Lindbladian contribution to the Liouville
superoperator.
For the one-qubit example, the only required
jump operator is
\begin{equation}
A=\left\vert 0 \right\rangle\left\langle 1\right\vert .
\end{equation}
Using this operator together with
Eq.~\eqref{eq:Gammafromgamma}, we obtain
\begin{equation}
\Gamma = \frac{\gamma }{2}\left(
\begin{array}{cccc}
 0 & 0 & 0 & 0 \\
 0 & 1 & i & 0 \\
 0 & -i & 1 & 0 \\
 0 & 0 & 0 & 0 \\
\end{array}
\right).
\end{equation}

Having obtained the quantities required to construct
the Hamiltonian and Lindbladian contributions to the
Liouville superoperator, namely \(\boldsymbol{H}\) and
\(\Gamma\), we can proceed to compute \(\Lambda\)
using Eq.~\eqref{eq:Lambdadef00}, yielding
\begin{equation}
\boldsymbol{\Lambda}
=\left(
\begin{array}{cccc}
 -\gamma & i \Omega  \cos (\Delta  t) & -i \Omega  \sin (\Delta  t) & \gamma/2 \\
 -i \Omega  \cos (\Delta  t) & \gamma/2 & i \gamma/2 & 0 \\
 i \Omega  \sin (\Delta  t) & -i \gamma/2 & \gamma/2 & 0 \\
 \gamma/2 & 0 & 0 & 0 \\
\end{array}
\right).
\end{equation}
The Liouville superoperator is obtained using
Eq.~\eqref{eq:liouvillefull01},
\begin{equation}
L=\left(
\begin{array}{cccc}
 0 & 0 & 0 & 0 \\
 0 & -\gamma/2 & 0 & \Omega  \sin (\Delta  t) \\
 0 & 0 & -\gamma/2 & \Omega  \cos (\Delta  t) \\
 \gamma & -\Omega  \sin (\Delta  t) & -\Omega  \cos (\Delta  t) & -\gamma \\
\end{array}
\right).
\end{equation}
As expected, all the elements of \(L\) are real.

In the Mathematica notebook \cite{lindbladalgebra},
the Liouville superoperator
is also computed using the direct method in the same
matrix basis for comparison.
The two methods are found to yield identical results.

In this example, we have considered a single-qubit
system, although the number of qubits can be increased.
The same calculations can be carried out for larger
systems ($n_q \ge 2$). Some of the computations,
however, involve high-rank tensors that,
for systems with more than two qubits,
may exceed the available memory.
The calculations most affected by this limitation
are indicated in the accompanying Mathematica notebook.

\section{Conclusions}\label{sec:conclusions}
We have determined the algebraic structure underlying the Liouville
master equation.
This structure has enabled us to express the Liouville master equation
in a compact and completely general form while also providing a
significant computational advantage over the direct approach.
In this formulation, the Liouville superoperator is written
as a linear combination of the elements of the operator set
$\mathfrak{X}_{n^2}$, which is closed under both the commutator
and the anticommutator.
Consequently, the dynamical map generated by the Liouville
superoperator belongs to the corresponding Lie group.

We have further shown that the Hermitian operators constituting
this algebra are mutually orthonormal with respect to the Hilbert-Schmidt
inner product. These results make it possible to derive explicit
differential equations governing the evolution of the coefficients
of the dynamical map.
Their solution yields the dynamical map $V(t)$
as a linear combination of the elements of the tensor $\mathfrak{X}_{n^2}$,
from which both the coherent vector and the density matrix can be
reconstructed directly.

An important consequence of this algebraic formulation
is that the inclusion of the Lindbladian fundamentally
enlarges the algebra governing the dynamics.
In the purely unitary case, corresponding to the
von Neumann equation, the evolution is generated solely
by the $n_1$ structure constants $C_i$.
In contrast, the general Lindblad equation requires the
full $n^2$-dimensional algebra $\mathfrak{X}_{n^2}$.
This shows that the Lindbladian does not merely modify
the coefficients of the evolution operator but fundamentally
enlarges the algebra required to represent the dynamics.

We have also derived a set of recursion relations that
enables the efficient construction of the algebra in
progressively larger dimensions.
Moreover, all model-dependent information is contained exclusively
in the tensor $\boldsymbol{\Lambda}$, and the Liouville superoperator
is obtained through a single tensor contraction.
These features substantially reduce the number of repeated
tensor products and contractions compared with the direct approach,
making the algebraic method computationally more efficient.
Furthermore, the bases $\boldsymbol{Z}$ and $\boldsymbol{X}$
are model independent.
Consequently, they can be precomputed and stored,
allowing them to be reused for any open quantum system
of a given dimension.

One potential application of this framework is the explicit
construction of the dynamical map,
from which the Choi matrix and the corresponding
Kraus operators can be obtained directly.

Finally, we have provided a link to a Mathematica notebook
containing a one-qubit example that illustrates the algebraic
solution of the dynamics of a transmon quantum circuit.
The same framework can be extended straightforwardly to
systems with larger numbers of qubits.

More generally, the present work establishes
a universal algebraic framework for finite-dimensional
Lindblad dynamics, in which the algebraic structure depends
only on the Hilbert-space dimension, while all physical information
is encoded in the tensor $\boldsymbol{\Lambda}$.
This separation between universal algebraic structure and model-dependent
dynamics provides both conceptual insight into the geometry of
open quantum systems and a practical foundation for efficient numerical
implementations.

\section{Acknowledgements}\label{sec:Acknowledgements}
GLA acknowledges the financial support of SECIHTI through
a Master's scholarship (CVU No.~2051081).
LB acknowledges the financial support of SECIHTI through
a PhD scholarship (CVU No.~960690).
AKB, VGIS, JCSS, and JLC acknowledge the financial support
of DCB UAM-A through grants 22322035 and 22322036.
VACB acknowledges the financial support of DCB UAM-A.

\appendix

\section{Recursion relations for the structure constants}
\label{app:a}

By employing the commutator and anticommutator
identities
\begin{eqnarray}
\left[A_2\otimes A_1, B_2\otimes B_1\right]
&=& \frac{1}{2}\left\{A_2,B_2\right\}\otimes \left[A_1,B_1\right]
+\frac{1}{2}\left[A_2,B_2\right]\otimes \left\{A_1,B_1\right\},\\
\left\{A_2\otimes A_1, B_2\otimes B_1\right\}
&=& \frac{1}{2}\left\{A_2,B_2\right\}\otimes \left\{A_1,B_1\right\}
+\frac{1}{2}\left[A_2,B_2\right]\otimes \left[A_1,B_1\right\},
\end{eqnarray}
one obtains recursion relations for the Lie and Jordan
structure constants \cite{Kunold_2024},
\begin{eqnarray}
c^{n_q+1}_{i,j,k}
&=&
\frac{1}{2}c^{n_q}_{i_2,j_2,k_2}b^{1}_{i_1,j_1,k_1}
+\frac{1}{2}b^{n_q}_{i_2,j_2,k_2}c^{1}_{i_1,j_1,k_1},
\\
b^{n_q+1}_{i,j,k}
&=&
\frac{1}{2}b^{n_q}_{i_2,j_2,k_2}b^{1}_{i_1,j_1,k_1}
-\frac{1}{2}c^{n_q}_{i_2,j_2,k_2}c^{1}_{i_1,j_1,k_1}.
\end{eqnarray}
Flattening indices, simplifies these expressions to
\begin{eqnarray}
C_k^{n_q+1} = \frac{1}{2}\left(C_{k_2}^{n_q}\otimes B_{k_1}^1
+ B_{k_2}^{n_q}\otimes C_{k_1}^1\right),\\
B_k^{n_q+1} = \frac{1}{2}\left(B_{k_2}^{n_q}\otimes B_{k_1}^1
- C_{k_2}^{n_q}\otimes C_{k_1}^1\right),
\end{eqnarray}
or, in tensorial notation,
\begin{eqnarray}
\boldsymbol{C}^{n_q+1}
= \frac{1}{2}\left(\boldsymbol{C}^{n_q}\otimes \boldsymbol{B}^1
+ \boldsymbol{B}^{n_q}\otimes \boldsymbol{C}^1\right),\\
\boldsymbol{B}^{n_q+1}
= \frac{1}{2}\left(\boldsymbol{B}^{n_q}\otimes \boldsymbol{B}^1
- \boldsymbol{C}^{n_q}\otimes \boldsymbol{C}^1\right).
\end{eqnarray}

\bibliography{paper-one.bib}

\end{document}